\documentclass[paper]{ieice}
\usepackage[pdftex]{graphicx,xcolor}
\usepackage[fleqn]{amsmath}
\usepackage{newtxtext}
\usepackage[varg]{newtxmath}
\usepackage{cite}
\usepackage{url}
\usepackage{diagbox}
\usepackage{multirow}
\usepackage[labelfont=bf]{caption}
\usepackage[font=footnotesize]{subcaption}
\field{B}

\title{Self-adaptive Traffic Anomaly Detection System\\for IoT Smart Home Environments}
\authorlist{%
 \authorentry[mf22145@shibaura-it.ac.jp]{Naoto Watanabe}{s}{sit}\MembershipNumber{2119371}
 \authorentry{Taku Yamazaki}{m}{sit}\MembershipNumber{1205099}
 \authorentry{Takumi Miyoshi}{m}{sit}\MembershipNumber{9411011}
 \authorentry{Ryo Yamamoto}{m}{uec}\MembershipNumber{0706639}
 \authorentry{Masataka Nakahara}{n}{kddir}\MembershipNumber{}
 \authorentry{Norihiro Okui}{m}{kddir}\MembershipNumber{1404121}
 \authorentry{Ayumu Kubota}{m}{kddir}\MembershipNumber{0715468}
}
\affiliate[sit]{The authors are with Graduate School of Engineering and Science, Shibaura Institute of Technology, Saitama-shi, 337-8570 Japan.}
\affiliate[uec]{The author is with Graduate School of Informatics and Engineering, The University of Electro-Communications, Chofu-shi, 182-8585 Japan.}
\affiliate[kddir]{The authors are with KDDI Research, Inc., Fujimino-shi, 356-8502 Japan.}

\received{2015}{1}{1}
\revised{2015}{1}{1}

\begin{document}
\maketitle

\begin{summary}
With the growth of internet of things (IoT) devices, cyberattacks, such as distributed denial of service, that exploit vulnerable devices infected with malware have increased.
Therefore, vendors and users must keep their device firmware updated to eliminate vulnerabilities and quickly handle unknown cyberattacks.
However, it is difficult for both vendors and users to continually keep the devices safe because vendors must provide updates quickly and the users must continuously manage the conditions of all deployed devices.
Therefore, to ensure security, it is necessary for a system to adapt autonomously to changes in cyberattacks.
In addition, it is important to consider network-side security that detects and filters anomalous traffic at the gateway to comprehensively protect those devices.
This paper proposes a self-adaptive anomaly detection system for IoT traffic, including unknown attacks.
The proposed system comprises a honeypot server and a gateway.
The honeypot server continuously captures traffic and adaptively generates an anomaly detection model using real-time captured traffic.
Thereafter, the gateway uses the generated model to detect anomalous traffic.
Thus, the proposed system can adapt to unknown attacks to reflect pattern changes in anomalous traffic based on real-time captured traffic.
Three experiments were conducted to evaluate the proposed system: a virtual experiment using pre-captured traffic from various regions across the world, a demonstration experiment using real-time captured traffic, and a virtual experiment using a public dataset containing the traffic generated by malware.
The results of all experiments showed that the detection model with the dynamic update method achieved higher accuracy for traffic anomaly detection than the pre-generated detection model.
The experimental results indicate that a system adaptable in real time to evolving cyberattacks is a novel approach for ensuring the comprehensive security of IoT devices against both known and unknown attacks.
\end{summary}


\section{Introduction}
\label{introduction}

The advent of the internet of things (IoT) era~\cite{iot-survey} has led to an increased use of various sensors and devices directly connected to the internet, which strive to improve our daily lives.
A smart home is a common example of an IoT-based system involving home automation, which includes remote monitoring and control of various smart home appliances that are IoT devices~\cite{iot-smarthome}.
Owing to their widespread use, the number of global active endpoints of IoT devices was reported to be 12.2 billion in 2021~\cite{2021iot}.
This number is expected to reach approximately 27 billion by 2025, indicating that the number of IoT devices will continue increasing.

With this growth of IoT devices, cyberattacks, such as distributed denial of service (DDoS) attacks, that exploit various vulnerable devices have also increased~\cite{ddos}.
In 2016, a botnet comprising IoT devices infected with the Mirai malware performed a targeted DDoS attack on a security blog~\cite{mirai-1,mirai-2}.
In 2021, a botnet comprising IoT devices infected with Meris attacked the KrebsOnSecurity website~\cite{meris}.

IoT devices infected with Mirai operate under default settings, including factory accounts, which allows the attacker to gain access through a brute-force attack.
Conversely, most devices infected with Meris were not updated with the latest firmware and remained vulnerable~\cite{meris}.
Thus, attackers continue to discover vulnerable devices and develop new variants of malware. Therefore, unknown cyberattacks are anticipated in the future.

To address such security issues, both the vendors and the users of IoT devices must maintain device security through continuous updates~\cite{iot-security-in-product-lifecycle,iot-firmware-secure-update-1,iot-firmware-secure-update-2}.
Vendors must continually track vulnerabilities and quickly update the device firmware to ensure security, and users must also manage and maintain the safety of their devices~\cite{iot-detecting-vulnerable-devices}.
However, it is still unrealistic to expect vendors to constantly provide up-to-date firmware for all IoT devices, including inexpensive ones~\cite{iot-firmware-update}.
Additionally, it is challenging for all users to maintain the safety of their devices as it might involve performing manual updates or applying appropriate settings.
Furthermore, IoT devices are designed to perform specific tasks with limited memory and computational resources for executing them.
Therefore, installing additional security modules, such as anti-malware software, within them is unrealistic.
Owing to these limitations, it is important to consider not only device safety but also an anomaly detection system that can adapt to unknown attacks from the network side.

This paper proposes a self-adaptive anomaly detection system for IoT traffic.
The proposed system adapts to unknown attacks by updating anomaly detection models using real-time captured traffic.
In addition, we developed the whole system as software modules working on Linux-based computers.
The proposed system comprises a honeypot server and a gateway.
The honeypot server captures traffic in real time, including cyberattacks and generates an anomaly detection model periodically from the captured traffic using machine learning.
The anomaly detection model classifies the traffic as malicious or benign based on the features of the input traffic.
The gateway monitors traffic and detects anomalies using the latest detection model.

The contributions of this study are as follows:
\begin{itemize}
    \item The proposed system adapts to changes in cyberattacks by applying a detection model using traffic captured in real time. Thus, this system has the potential to address unknown cyberattacks whose attack methods have not been elucidated.
    \item We developed an easy-to-use traffic anomaly detection system as software modules working on Linux-based computers. This system comprises three software modules: traffic capturing module, model update module, and anomaly detection module implemented in a gateway and a honeypot server.
    \item The proposed system is flexible because the operator can select their preferred honeypot and machine learning algorithm.
    \item We evaluated the proposed system using pre-captured traffic in worldwide, real-time traffic in a real smart home environment, and a public dataset. We confirmed that dynamically updated detection models can detect anomalous traffic with higher accuracy than pre-generated detection models.
\end{itemize}

The remainder of this paper is organized as follows:
Section \ref{related} presents related work on IoT device security. 
Section \ref{proposed} introduces the proposed self-adaptive anomaly detection system for IoT traffic. 
Section \ref{experimental} elucidates the experiments performed for evaluating the proposed system.
Section \ref{result} presents the experimental results of the proposed system for real smart-home traffic. 
Section \ref{conclusion} concludes the paper and provides insights for future work.

\begin{figure*}[t]
	\centering
	\includegraphics[clip, width=0.95\linewidth]{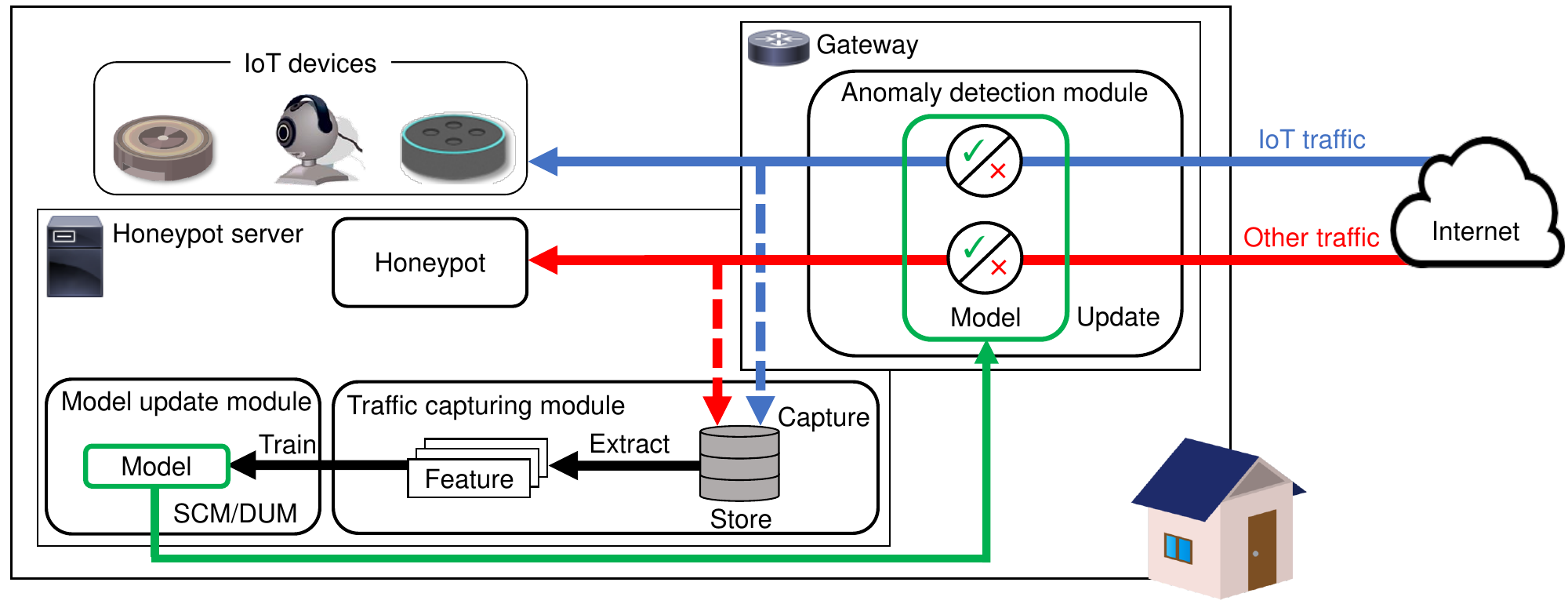}
	\caption{Overview of the proposed system. The proposed system is composed of the gateway, honeypot server, and IoT devices. The gateway observes the traffic and detects traffic anomalies using a machine learning model. The honeypot server trains the machine learning model using the captured traffic passed through the gateway and provides it to the gateway.}
	\label{zu1}
\end{figure*}

\section{Related Work}
\label{related}

Recently, many studies have focused on threat detection to protect IoT devices~\cite{anomaly-detection-survey,mts1,mts2,anomaly-detection-using-anomalous-data,hung,anomaly-based,ad-iot,unsw,ultra,neural,consume,entropy,edima,unsupervised}.
We categorized them into four types: general anomaly detection~\cite{anomaly-detection-survey,mts1,mts2,anomaly-detection-using-anomalous-data,hung,anomaly-based}, general intrusion and attack detection without considering a specific attack~\cite{ad-iot,unsw,ultra}, DDoS attack detection~\cite{neural,consume,entropy}, and early-stage scanning activity detection~\cite{edima,unsupervised}.

\noindent\textbf{General anomaly detection.}
Many anomaly detection methods in IoT environments using statistical and machine learning methods were proposed~\cite{anomaly-detection-survey}.
Generally, dimensionality reduction-based methods were proposed to detect abnormal behavior within multivariate time-series data by comparing with the normal behavior~\cite{mts1,mts2}.
In contrast, a method using a dataset containing anomalous data was proposed to detect anomalies by learning abnormal behaviors~\cite{anomaly-detection-using-anomalous-data}.
In either case, these methods need to define normal or abnormal behaviors to establish a baseline for detecting anomalies.
However, methods based on abnormal device behaviors are difficult to detect unknown attacks because these attacks are not included in the learned model.
On the other hand, methods that define normal device behaviors are expected to detect unknown attacks because they identify deviations from the established baseline of normality as anomalies.
However, the normal behavior of some devices, such as smart speakers, are difficult to define in smart home environments because these devices generate user-dependent irregular traffic~\cite{hung}.
Therefore, anomaly detection methods increase false positives as the device behavior diversifies~\cite{anomaly-based}.

\noindent\textbf{General intrusion and attack detection.} 
In~\cite{ad-iot}, the authors proposed an intrusion detection system that uses machine learning to detect various cyberattacks within IoT traffic. 
They evaluated the system on the UNSW-NB15 dataset~\cite{unsw}, which contains pre-captured traffic, including normal and abnormal traffic, as the training data.
In~\cite{ultra}, the authors proposed an attack detection method using bit pattern matching of packet payloads, which can be executed on IoT devices owing to its lightweight algorithm.
They focused on detecting known attacks, and their method requires pre-generating matching patterns from pre-captured traffic information.
Therefore, this method might not be suitable for handling future unknown attacks.

\noindent\textbf{DDoS detection.} 
In~\cite{neural}, the authors created a dataset comprising the traffic generated by IoT devices in an urban environment and traffic-emulated DDoS attacks.
They used the created dataset and deep-learning algorithms to detect DDoS attacks.
In~\cite{consume}, the authors proposed a method to detect DDoS attacks on the gateway through machine learning by using the traffic of actual attacks performed by Mirai-infected IoT devices.
They used the source code of Mirai, which is publicly available on the internet, to attack the experimental servers and capture traffic.
In~\cite{entropy}, the authors proposed an entropy-based method to detect DDoS attacks performed by malware-infected IoT devices on a gateway.
This method calculates the entropy of each device and determines whether a device behaves as a DDoS attacker by checking if the entropy exceeds a threshold value.
These studies mainly focused on detecting DDoS attacks. 
Therefore, they cannot detect traffic anomalies other than DDoS attacks.

\noindent\textbf{Scanning activity detection.}
In~\cite{edima}, the authors proposed a method to detect scanning activities of malware-infected IoT devices, which are the early stages of attacks, by using machine learning with few features.
The authors of~\cite{unsupervised} proposed a method for detecting scanning activities of malware-infected IoT devices using unsupervised machine learning.
These studies focused on scanning activities and did not consider other abnormal activities. 
Additionally, these methods cannot adapt to some activities in the local environment because the quality of the learned model depends on local activity. 

In summary, previous studies mainly focused on specific attacks or improving attack detection accuracy using pre-captured traffic.
In other words, they did not focus on the response to temporal changes in device behavior and cyberattacks.
However, IoT environments, such as smart homes, require continuous and self-adaptive device protection to mitigate unknown attacks, as explained in Sec.~\ref{introduction}.

\section{Self-adaptive Traffic Anomaly Detection System for IoT Smart Home Environments}
\label{proposed}

This paper proposes a self-adaptive anomaly detection system for IoT traffic in smart home environments.
The proposed system adopts a honeypot server and machine learning to collect real-time traffic information and autonomously adapt to and detect unknown cyberattacks based on the collected information.

Fig.~\ref{zu1} shows an overview of the proposed system comprising two nodes: a gateway that detects attacks using an anomaly detection model in addition to operating as a general commercial router, and a honeypot server that continuously captures real-time traffic, including attacks, and generates an anomaly detection model using the captured traffic. 
The anomaly detection model is a binary classification model that classifies malicious or benign hosts based on the input traffic features.

Additionally, the proposed system comprises three software modules developed in Python that work on Linux-based computers: 
(1) traffic capturing module, (2) model update module, and (3) anomaly detection module. 
The traffic capturing and model update modules run on the honeypot server, whereas the anomaly detection module runs on the gateway, as illustrated in Fig.~\ref{zu1}.

The proposed system operates as follows:
First, the traffic capturing module continuously captures all incoming and outgoing traffic in real-time, including attacks, using a honeypot exposed to the internet. 
It should be noted that the gateway permits direct access to the honeypot from the internet, and vice versa, and mirrors the traffic generated by the local IoT devices to the honeypot server.

Second, the model update module dynamically executes the training phase using stored traffic to generate the latest detection model. 
Thereafter, either of the two model update methods is employed, static creation method (SCM) or dynamic update method (DUM), as explained in Sec.~\ref{proposed:update}. 
If the operator selects DUM, the update process is executed periodically using the latest real-time captured traffic to mitigate unknown attacks.
During the model update process, the model update module extracts features from the stored traffic and executes a machine learning algorithm to train a detection model. 
The detailed algorithm adopted in this study is described in \ref{proposed:feature}.

After generating the detection model, the honeypot server sends it to the gateway. 
When the anomaly detection module on the gateway receives the latest model from the honeypot server, the gateway replaces the existing model with the received model. 
Thereafter, it uses the latest model to detect anomalies in the passing traffic.

The proposed system periodically repeats these processes to autonomously adapt to unknown attacks by updating the anomaly detection model based on the latest real-time traffic.
In most cases, abnormal traffic can be collected irregularly because malicious traffic such as cyberattacks occurs infrequently. 
An imbalance between normal and abnormal traffic causes performance degradation as a common problem in machine learning~\cite{unbalanced1,unbalanced2}. 
In contrast, the proposed system can continuously collect the latest cyberattack traffic by using a honeypot. 
In other words, the proposed system can resolve the problem caused by imbalanced data by appropriately adjusting the balance between normal and abnormal traffic for machine learning. 
From a practical perspective, the proposed system is flexible because it is composed of multiple software modules. 
Therefore, operators can run the proposed system with the preferred honeypot and gateway. 
In addition, they can select an anomaly detection algorithm to suit for their environments.

\begin{table}[t!] 
  \centering
  \caption{Honeypots running on the honeypot server used in the proposed system. Specific attacks are captured by corresponding honeypot services. The others are captured by Honeytrap.}
  \label{tab:tpot}
  \begin{tabular}{ccc}
    \hline
      Name                 & Protocol & Observed port \\ \hline
      ADBHoney~\cite{adbhoney} & TCP & 5555 \\
      CitrixHoneypot~\cite{citrixhoneypot} & TCP & 443 \\      
      Conpot~\cite{conpot} & TCP & 1025, 2404, 10001, 50100 \\
      Conpot~\cite{conpot} & UDP & 161, 623 \\
      Cowrie~\cite{cowrie} & TCP & 22, 23 \\
      Dicompot~\cite{dicompot} & TCP & 11112 \\
      Dionaea~\cite{dionaea} & TCP & 21, 42, 81, 135, 445, \\
                 &        & 1433, 1723, 1883, 3306, \\ 
                 &        & 5061, 27017 \\
      Dionaea~\cite{dionaea} & UDP & 69, 5060 \\
      ElasticPot~\cite{elasticpot} & TCP & 9200 \\
      Heralding~\cite{heralding} & TCP & 110, 143, 465, 993, 995, \\ 
                   &       & 1080, 5432, 5900 \\
      HoneySAP~\cite{honeysap} & TCP & 3299 \\
      Mailoney~\cite{mailoney} & TCP & 25 \\
      Medpot~\cite{medpot} & TCP & 2575 \\
      RDPY~\cite{rdpy} & TCP & 3389 \\
      Snare~\cite{snaretanner} & TCP & 80 \\
      Tanner~\cite{snaretanner} & TCP & 6379 \\
      Honeytrap~\cite{honeytrap} & TCP/UDP & others \\    
    \hline
  \end{tabular}
\end{table}

\subsection{Honeypot Server and Traffic Capturing}
\label{proposed:capture}

The honeypot server continuously captures and stores the incoming and outgoing traffic from the internet using the traffic capturing module.
The honeypot server operates a honeypot to observe and record the behavior of attacks from the internet. 
The operator can select their preferred honeypot.

In this study, we adopted the T-Pot~\cite{tpot} platform as the honeypot.
T-Pot is a multi-honeypot platform based on Debian 11.
T-Pot allows multiple honeypots to run on independent Docker containers and receive various attacks through a single network interface.
Table~\ref{tab:tpot} lists the protocols and ports of the honeypots running on the T-Pot platform~\cite{adbhoney,citrixhoneypot,conpot,cowrie,dicompot,dionaea,elasticpot,heralding,honeysap,mailoney,medpot,rdpy,snaretanner,honeytrap} employed in our honeypot server. 
The Honeytrap~\cite{honeytrap} honeypot observes the ports that are not observed by the other honeypots.

\subsection{Methods for Updating the Anomaly Detection Model}
\label{proposed:update}

As mentioned previously, the honeypot server continuously captures and stores real-time traffic. 
This subsection explains the methods used for updating the anomaly detection model.

\begin{figure}[t]
	\centering
	\includegraphics[clip, width=0.95\linewidth]{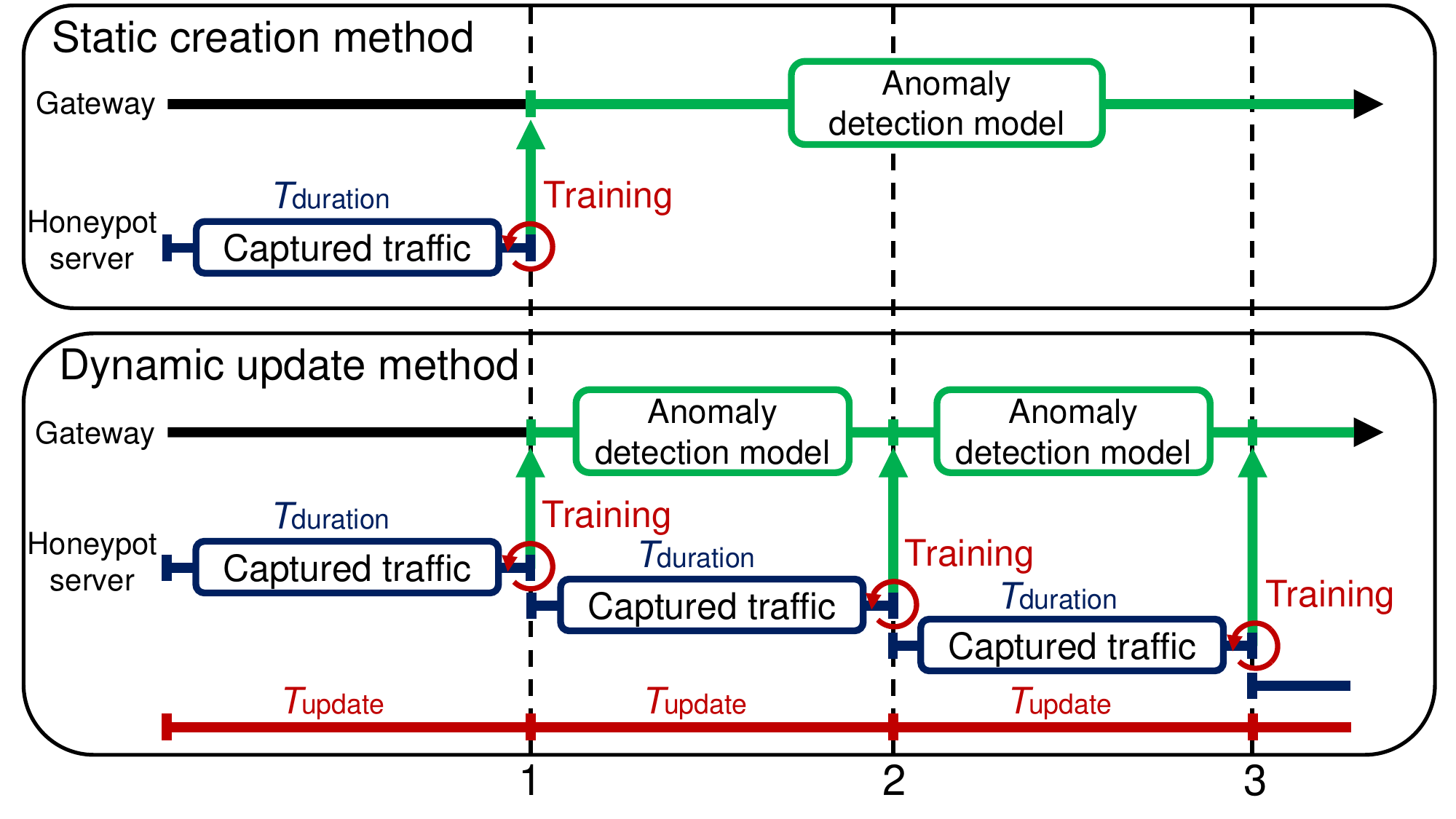}
    \caption{Illustrations of the operation of the static creation method ($T_{\mathrm{duration}}=1$ [h]) and dynamic update method ($T_{\mathrm{duration}}, T_{\mathrm{update}}=1$ [h]).}
	\label{zu2}
\end{figure}

The system operator can select either SCM or DUM as the update method.
Examples of the operations included in each method are shown in Fig.~\ref{zu2}. 
$\mathcal{D}$ denotes the dataset of the entire traffic captured by the honeypot server and $\mathcal{D}_{t-1,t}$ denotes a partial dataset of $\mathcal{D}$ from previous time $t-1$ to current time $t$. 
This partial dataset is used to create and update the detection model.
$T_{\mathrm{duration}}$ denotes the duration of traffic capture and $T_{\mathrm{update}}$ denotes the update interval of the detection model.

\noindent\textbf{Static creation method (SCM)}. In SCM, the honeypot server captures and stores the traffic~$\mathcal{D}_{0,T_{\mathrm{duration}}}$ from the system start time ($t=0$) until $T_{\mathrm{duration}}$. 
Subsequently, it creates an anomaly detection model using the captured traffic~$\mathcal{D}_{0,T_{\mathrm{duration}}}$. 
Next, the honeypot server sends the created model to the gateway, which uses it for anomalous traffic detection.
When using SCM, the proposed system need not periodically update the model on the gateway and the honeypot server need not continuously capture or learn the traffic. 
Therefore, although the anomaly detection model can be realized with few computational and network resources, its quality strongly depends on the initial dataset.

\noindent\textbf{Dynamic update method (DUM)}. In DUM, if the current time is $t$, the honeypot server always captures and stores traffic~$\mathcal{D}_{t-T_{\mathrm{duration}},t}$ from $t-T_{\mathrm{duration}}$ to $t$. 
It periodically trains the selected machine learning algorithm with interval~$T_{\mathrm{update}}$ using the captured traffic~$\mathcal{D}_{t-T_{\mathrm{duration}},t}$ and sends the created detection model to the gateway. 
Thereafter, the gateway replaces the previous model with the received model and uses the updated model for anomalous traffic detection.
In DUM, the detection model can be adaptively updated at $T_{\mathrm{update}}$ intervals.
Therefore, DUM can handle the changes in attacks in real-time to detect new unknown anomalous traffic. 
Additionally, it can reduce the volume of stored traffic because it at most stores only partial traffic from duration~$T_{\mathrm{duration}}$ to current time~$t$. 
However, it may degrade the quality of the detection model compared to the ideal case of using all the captured traffic.

\subsection{Feature Extraction and Machine Learning}
\label{proposed:feature}

This subsection explains the method of feature extraction from the captured traffic in the proposed system.

\begin{figure}[t]
	\centering
	\includegraphics[clip, width=0.95\linewidth]{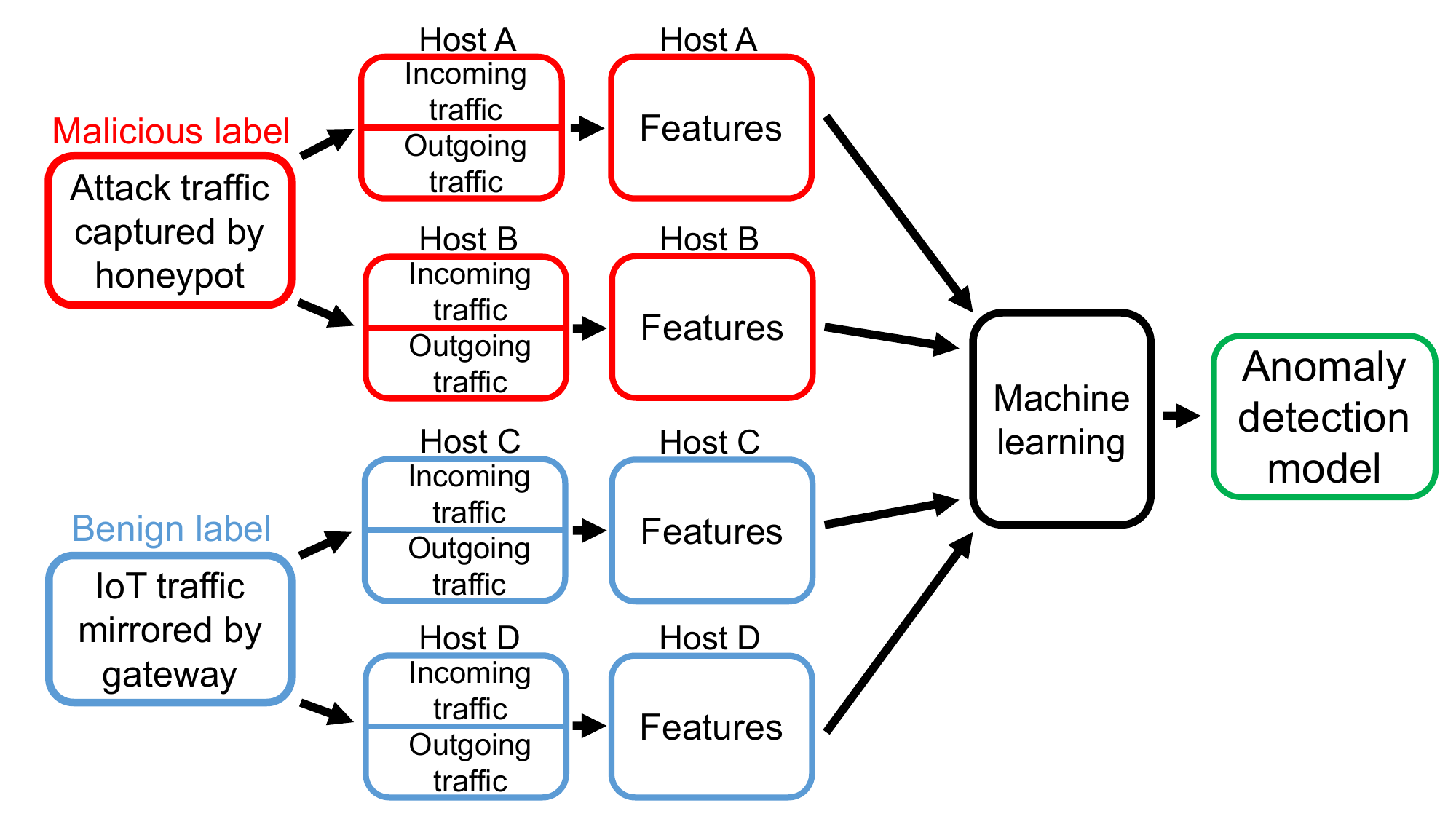}
	\caption{Procedure for creating an anomaly detection model.}
	\label{zu3}
\end{figure}

Fig.~\ref{zu3} shows an example of feature extraction using real-time captured traffic in the proposed system.
First, the honeypot server divides the traffic captured for each host and labels the hosts communicating with local IoT devices and the honeypot server as benign and malicious, respectively.
Next, it segregates the traffic according to each host into incoming and outgoing traffic and extracts features from the traffic generated by communication with each host.
Table~\ref{tab:feat} lists the features used in the proposed system, which are calculated using the header information of each traffic packet.

Finally, the honeypot server creates a binary classification model of benign or malicious using the extracted features through supervised learning. 
The operator can select the optimal machine learning algorithm for model creation.

\begin{table}[t]
  \centering
  \caption{Extracted features used in our experiments.}
  \label{tab:feat}
  \scalebox{0.97}[1]{ 
    \begin{tabular}{cc}
      \hline
      Incoming traffic & Outgoing traffic \\ \hline
      Minimum packet receiving interval & Minimum packet sending interval \\
      Maximum packet length & Maximum packet length \\
      Minimum packet length & Minimum packet length \\
      Protocol &  \\
      Destination port &  \\
      Time to live &  \\   
      \hline
    \end{tabular}
  }
\end{table}

\subsection{Anomaly Detection on Gateway}
\label{proposed:detection}

After the honeypot server generates an anomaly detection model, it sends it to the gateway, which then replaces the existing model with the received model. 
Thereafter, the gateway uses the model to detect anomalies in the real-time observed traffic. 
The gateway determines whether a host is benign or malicious based on observed traffic features. 
If a host is determined as malicious, the gateway records its IP address in its IP address list of malicious hosts.

Note that the current implementation includes two types of traffic anomaly detection behaviors: (1) filtering and discarding the anomalous traffic generated by malicious hosts on the gateway and (2) recording the traffic as anomalous and passing the gateway to the honeypot server. 
We adopted the mechanism of passing the traffic to the honeypot server.

\begin{table}[t]
  \centering
  \caption{Experiment 1: IoT devices virtually installed in the smart home environment.}
  \label{tab:iot-exp1}
  \begin{tabular}{ccc}
    \hline
      \multirow{2}{*}{Category} & \multirow{2}{*}{Device name} & Number of \\ 
                                         &                     &  devices  \\ \hline
      Smart Scale & Elecom Eclear & 1 \\
      Smart Plug & Meross Smart Power Strip & 1 \\
      Smart Plug & Meross Smart Wifi Plug Mini & 1 \\
      Humidifier & Meross Humidifier & 1 \\
      Smart Bulb & Meross Smart Wifi LED Bulb & 1 \\
      Remote Controller & Meross Smart IR Remote Control & 1 \\
      Smart Speaker & Google Nest Mini & 1 \\
      Network Camera & ATOM Tech Atom Cam & 1 \\
      Smart Bulb & TP-Link Kasa Smart LED Bulb & 1 \\
      Network Camera & TP-Link Kasa Pro & 1 \\
      Tablet & Amazon Fire 7 Tablet & 1 \\
      Smart Speaker & Amazon Echo Dot with Clock & 1 \\
      Smart Hub & Panasonic Home Unit & 1 \\
      Sensor & Panasonic Motion Sensor & 1 \\
      Network Camera & Panasonic Network Camera & 1 \\
      Sensor & LinkJapan eSensor & 1 \\      
      Robot Cleaner & Dyson 360 Heurist & 1 \\ 
    \hline
  \end{tabular}
\end{table}

\begin{figure*}[t!]
	\centering
	\includegraphics[clip, width=0.95\linewidth]{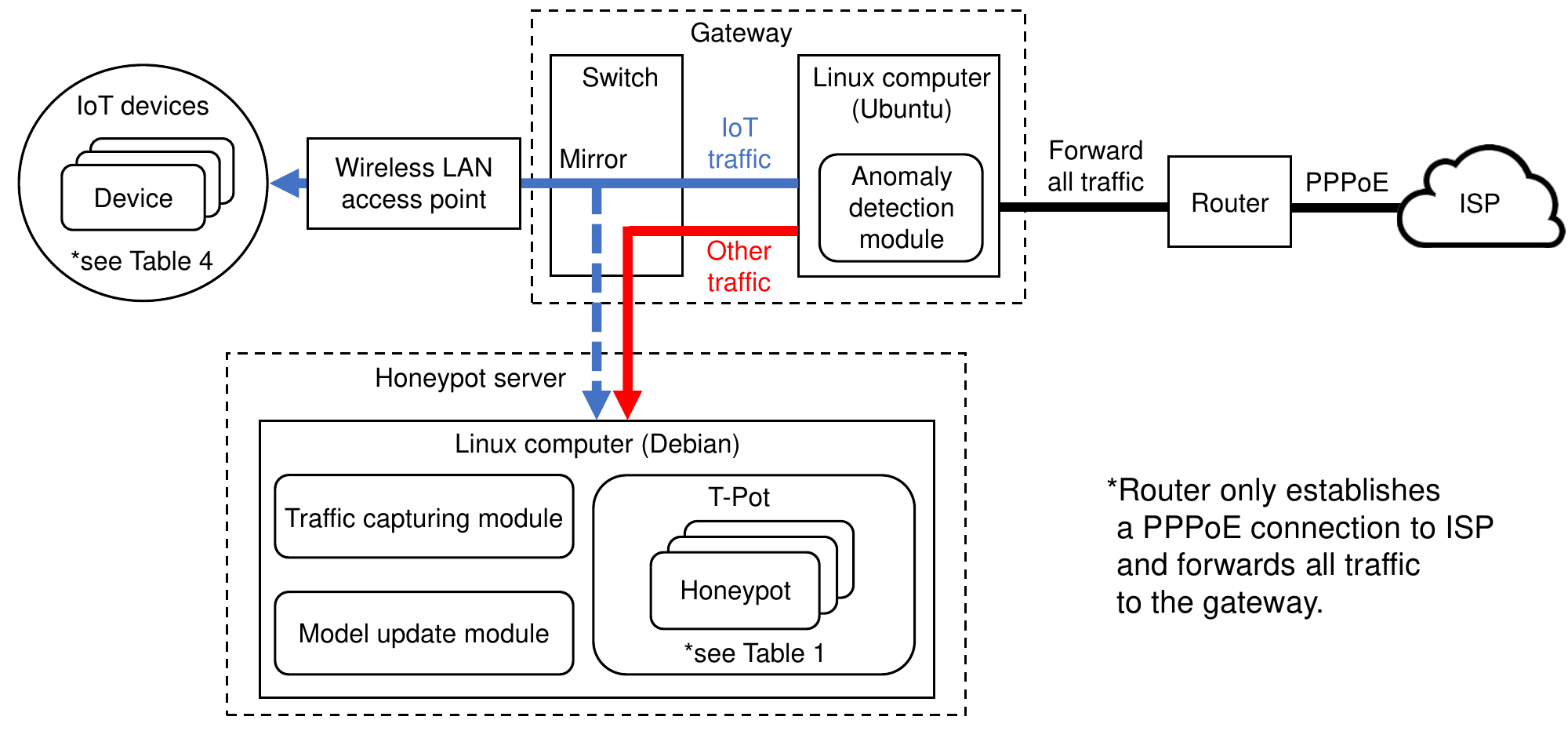}
	\caption{Experiment 2: Overview of the smart home environment used in the experiment.}
	\label{zu4}
\end{figure*}

\section{Experimental Environments}
\label{experimental}

\subsection{Experiment Setup}

To demonstrate the effectiveness of the proposed system, we conducted three experiments to detect anomalous traffic: 
(1) a virtual experiment using pre-captured traffic from multiple regions across the world, (2) a demonstration experiment using real-time captured traffic in a real smart home environment, and (3) a virtual experiment using a public dataset containing the traffic generated by malware.

\noindent\textbf{Experiment 1}: To demonstrate the performance and effectiveness of the proposed system in various regions across the world, we conducted an experiment using pre-captured traffic from different locations on the Microsoft Azure platform~\cite{azure} in addition to the pre-captured traffic of IoT devices in our laboratory. 
Virtual servers with pre-installed T-Pot were deployed in the regions of Australia East, US East, and Japan East and these were exposed to the internet to capture malicious traffic.
Additionally, we parallelly captured the normal traffic generated by the commercial IoT devices listed in Table~\ref{tab:iot-exp1}. 
In our laboratory, we prepared an actual smart home environment comprising various IoT devices that were connected to the internet through a wireless access point. 
Raspberry Pi was employed as the traffic capturing device, and the traffic passing through the access point was mirrored onto it.
All traffic was captured in parallel over one month, 1--31 January, 2022 (JST).

\noindent\textbf{Experiment 2}: To confirm the effectiveness of the proposed system in a real environment, we conducted a demonstration experiment in an actual smart home environment comprising a home network and commercial IoT devices. 
Fig.~\ref{zu4} shows an overview of Experiment 2.
We prepared a typical home network in our laboratory, which was directly connected to the internet via a fiber optic broadband service typically used in Japanese households. 
The network was completely isolated from our university network by installing an exclusive line to reproduce a home environment. 
A commodity router was used to establish the internet connection to Plala~\cite{plala} as the internet service provider. 
Under the commodity router, we connected a Linux-based computer (Ubuntu 20.04.6 LTS) as a gateway to run an anomaly detection module.
The commodity router forwarded all traffic to the gateway to reproduce an environment wherein the gateway was directly connected and exposed to the internet. 
We connected the T-Pot and access point with IoT devices to the network switch installed under the gateway and configured the switch to mirror traffic passing through the access point to the T-Pot.
Furthermore, the gateway allowed direct access from the internet to the T-Pot and captured traffic on it.
Table~\ref{tab:iot-exp2} lists the IoT devices installed in the experimental smart home environment. 
Note that SwitchBot Hub Mini includes a function for managing devices and a programmable remote controller for non-smart devices. 
We connected three door sensors, one smart switch, one smart lock, and four thermo-hygrometers to two hub devices through Bluetooth. 
The experiment was conducted over two weeks, from 18:00 on the 3rd to 18:00 on the 17th of August, 2022 (JST).

\noindent\textbf{Experiment 3}: To confirm the effectiveness of the model update in a mixed environment of normal and malware-infected abnormal devices, we conducted an experiment using a public dataset containing malicious traffic generated by malware. 
For abnormal traffic, we used the IoT-23 dataset~\cite{iot23}, which contains traffic generated by Mirai~\cite{mirai-1,mirai-2}, Hajime~\cite{hajime}, and Hakai~\cite{hakai} running on Raspberry Pi, captured for 11 days, 21-31 July, 2018 (JST).
For normal traffic, we used the traffic generated by the normal behavior of IoT devices captured by Experiment 1, 1-11 January, 2022 (JST).

\begin{table}[t]
  \centering
  \caption{Experiment 2: IoT devices installed in the smart home environment.}
  \label{tab:iot-exp2}
  \begin{tabular}{ccc}
    \hline
      \multirow{2}{*}{Category} & \multirow{2}{*}{Device name} & Number of \\ 
                                         &                     &  devices  \\ \hline
      Smart Hub & SwitchBot Hub Mini     & 2 \\
      Smart Bulb & SwitchBot Color Bulb   & 4 \\
      Tape Light & SwitchBot LED Strip Light & 2 \\
      Air Purifier & SHARP KI-JS70-H      & 1 \\
      Smart Speaker & Amazon Echo Show 5 & 3 \\     
    \hline
  \end{tabular}
\end{table}

\begin{figure*}[t!]
  \centering
  \begin{tabular}{ccc}
    \begin{minipage}{0.3\hsize}
      \centering
      \includegraphics[clip, width=1.0\linewidth]{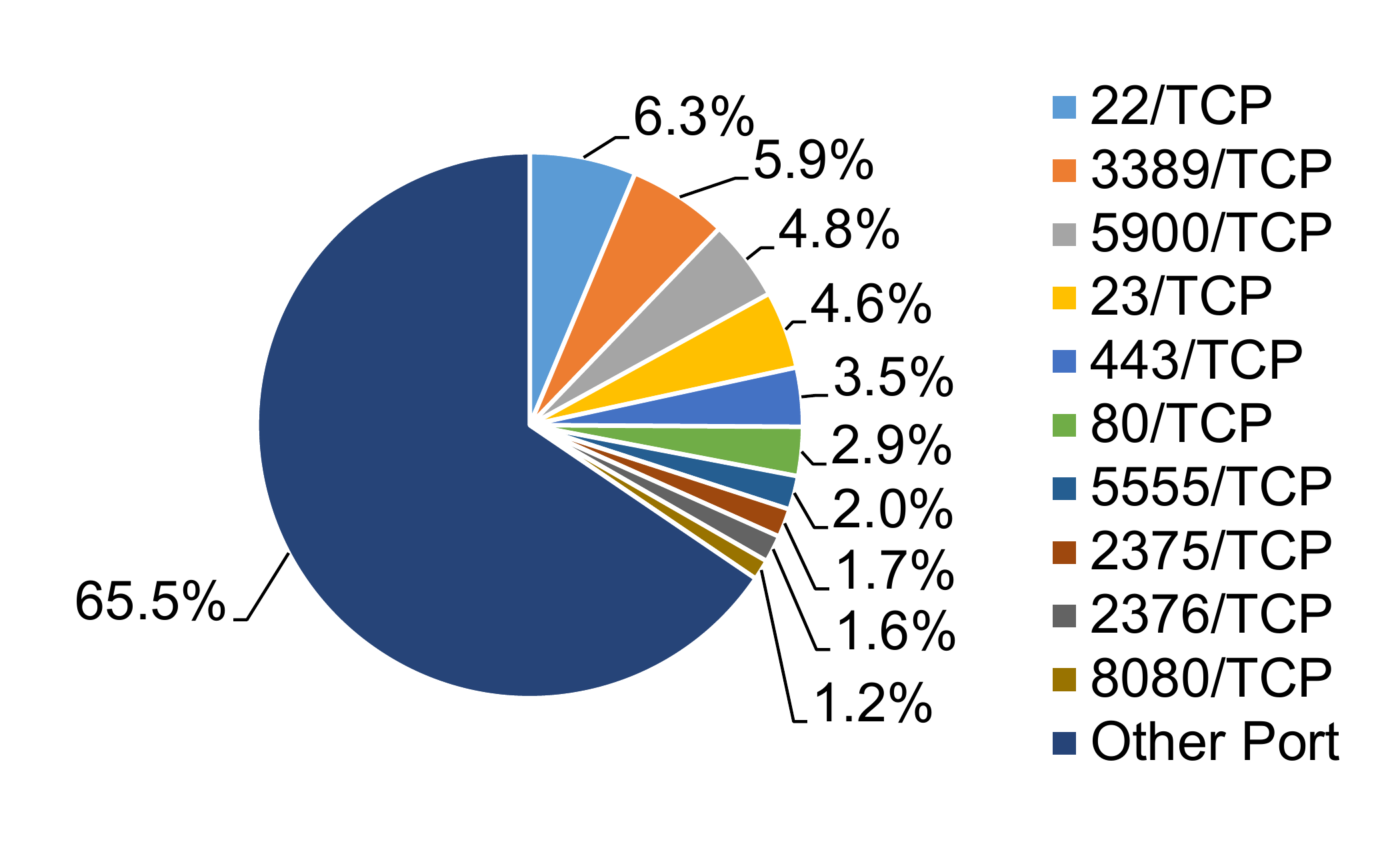}
      \subcaption{Australia East}
    \end{minipage}
    \begin{minipage}{0.3\hsize}
      \centering
      \includegraphics[clip, width=1.0\linewidth]{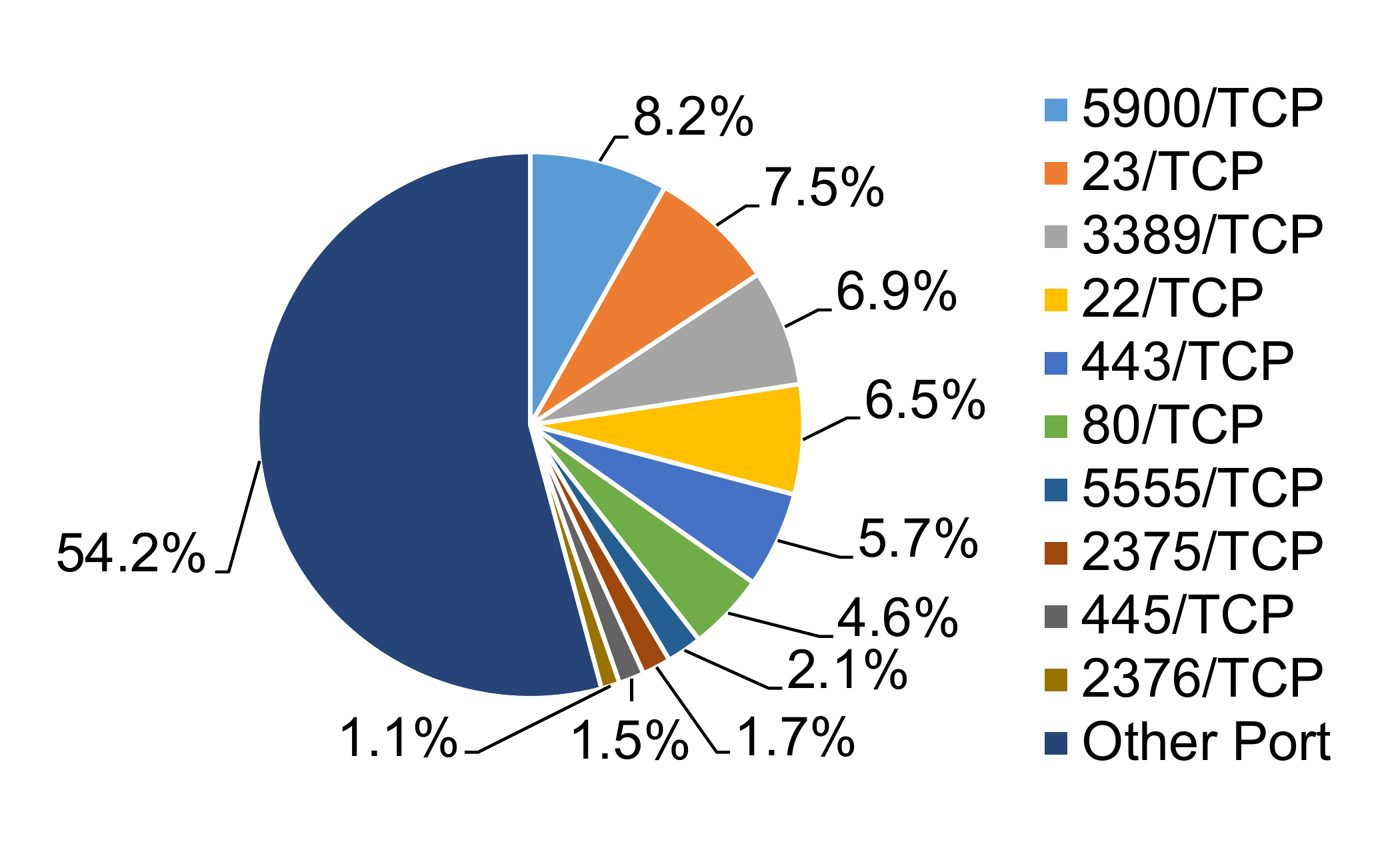}
      \subcaption{US East}
    \end{minipage}
    \begin{minipage}{0.3\hsize}
      \centering
      \includegraphics[clip, width=1.0\linewidth]{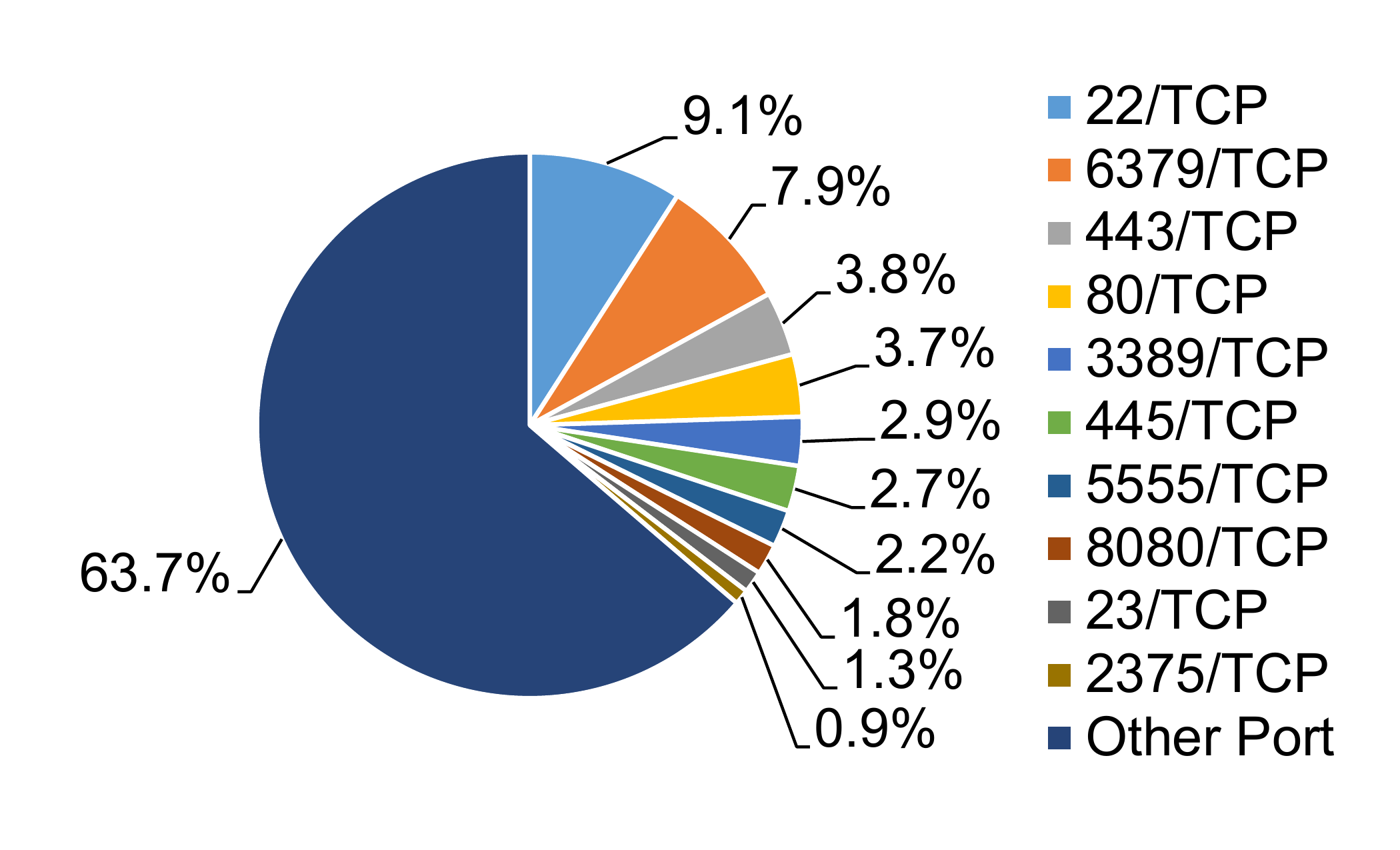}
      \subcaption{Japan East}
    \end{minipage}
  \end{tabular}
  \caption{Experiment 1: Distributions of destination ports observed by the honeypot server.}
  \label{dstport-azure}
\end{figure*}

\begin{figure*}[t!]
  \centering
  \begin{tabular}{cccccc}
    \begin{minipage}{0.16\hsize}
      \centering
      \captionsetup{justification=centering}
      \includegraphics[clip, width=\hsize]{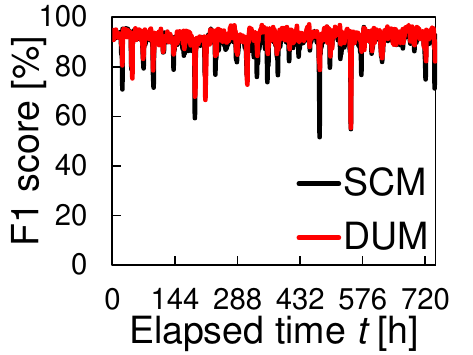}
      \subcaption{Australia East\\(SVC)}
    \end{minipage}
    \begin{minipage}{0.16\hsize}
      \centering
      \captionsetup{justification=centering}
      \includegraphics[clip, width=\hsize]{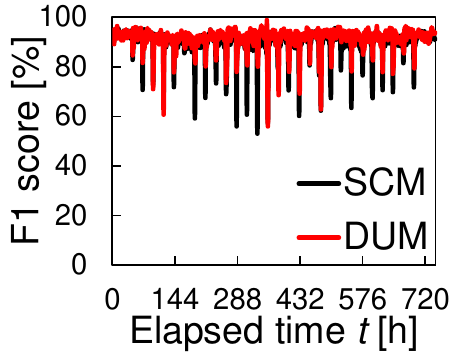}
      \subcaption{US East\\(SVC)}
    \end{minipage}
    \begin{minipage}{0.16\hsize}
      \centering
      \captionsetup{justification=centering}
      \includegraphics[clip, width=\hsize]{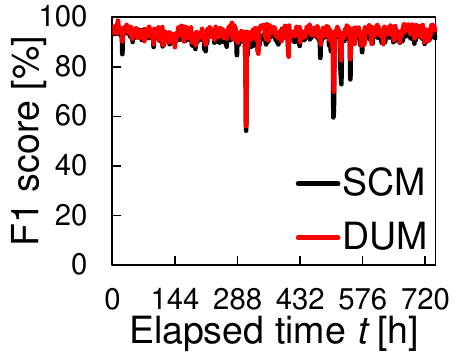}
      \subcaption{Japan East\\(SVC)}
    \end{minipage}
    \begin{minipage}{0.16\hsize}
      \centering
      \captionsetup{justification=centering}
      \includegraphics[clip, width=\hsize]{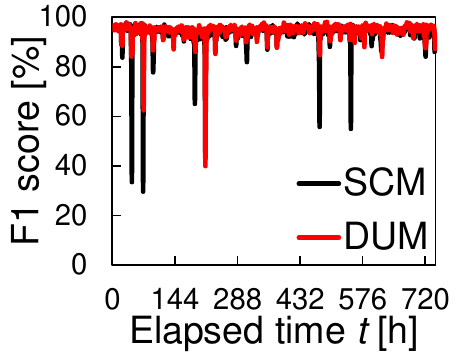}
      \subcaption{Australia East\\(k-NN)}
    \end{minipage}
    \begin{minipage}{0.16\hsize}
      \centering
      \captionsetup{justification=centering}
      \includegraphics[clip, width=\hsize]{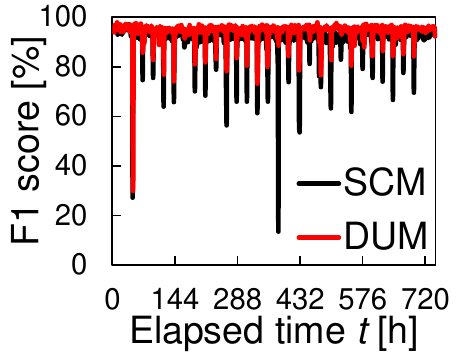}
      \subcaption{US East\\(k-NN)}
    \end{minipage}
    \begin{minipage}{0.16\hsize}
      \centering
      \captionsetup{justification=centering}
      \includegraphics[clip, width=\hsize]{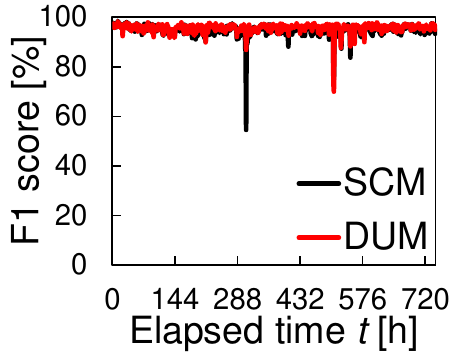}
      \subcaption{Japan East\\(k-NN)}
    \end{minipage}
    \vspace{1mm}
    \\
    \begin{minipage}{0.16\hsize}
      \centering
      \captionsetup{justification=centering}
      \includegraphics[clip, width=\hsize]{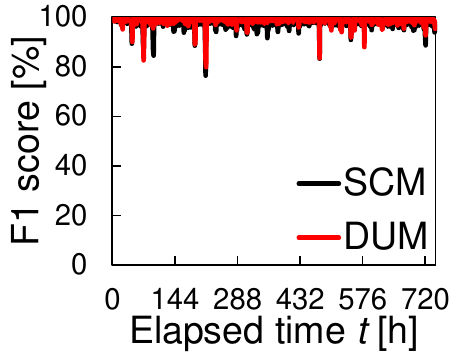}
      \subcaption{Australia East\\(DT)}
    \end{minipage}
    \begin{minipage}{0.16\hsize}
      \centering
      \captionsetup{justification=centering}
      \includegraphics[clip, width=\hsize]{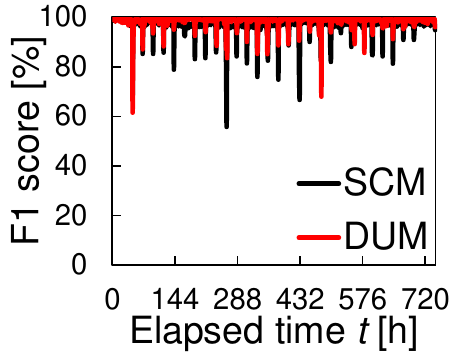}
      \subcaption{US East\\(DT)}
    \end{minipage}
    \begin{minipage}{0.16\hsize}
      \centering
      \captionsetup{justification=centering}
      \includegraphics[clip, width=\hsize]{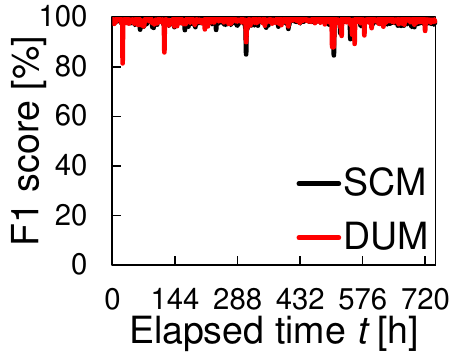}
      \subcaption{Japan East\\(DT)}
    \end{minipage}
    \begin{minipage}{0.16\hsize}
      \centering
      \captionsetup{justification=centering}
      \includegraphics[clip, width=\hsize]{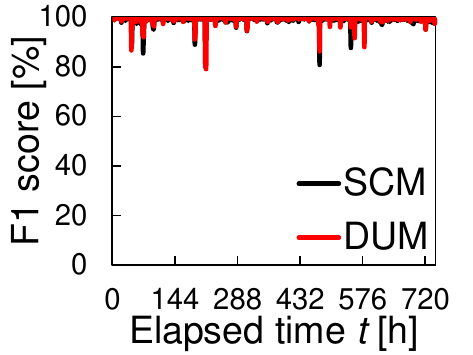}
      \subcaption{Australia East\\(RF)}
    \end{minipage}
    \begin{minipage}{0.16\hsize}
      \centering
      \captionsetup{justification=centering}
      \includegraphics[clip, width=\hsize]{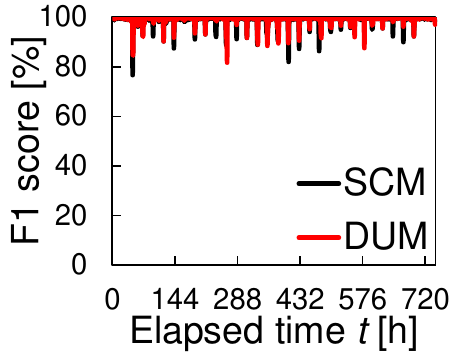}
      \subcaption{US East\\(RF)}
    \end{minipage}
    \begin{minipage}{0.16\hsize}
      \centering
      \captionsetup{justification=centering}
      \includegraphics[clip, width=\hsize]{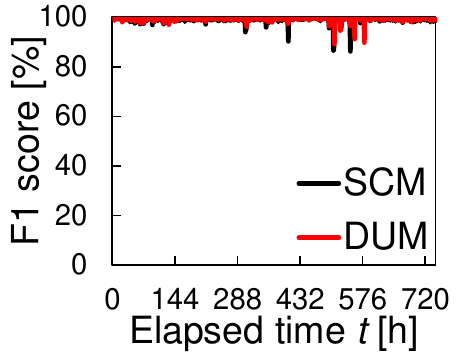}
      \subcaption{Japan East\\(RF)}
    \end{minipage}
    \vspace{1mm}
    \\
    \begin{minipage}{0.16\hsize}
      \centering
      \captionsetup{justification=centering}
      \includegraphics[clip, width=\hsize]{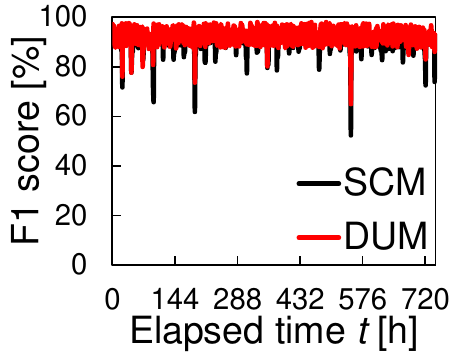}
      \subcaption{Australia East\\(ANN)}
    \end{minipage}
    \begin{minipage}{0.16\hsize}
      \centering
      \captionsetup{justification=centering}
      \includegraphics[clip, width=\hsize]{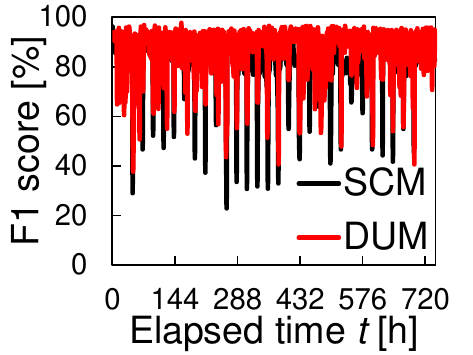}
      \subcaption{US East\\(ANN)}
    \end{minipage}
    \begin{minipage}{0.16\hsize}
      \centering
      \captionsetup{justification=centering}
      \includegraphics[clip, width=\hsize]{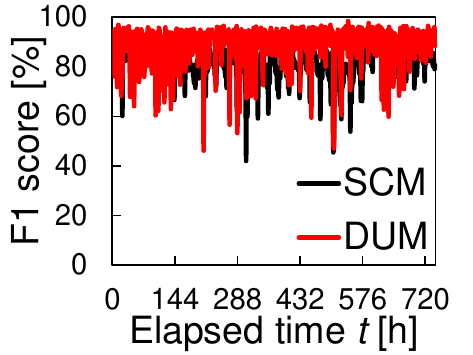}
      \subcaption{Japan East\\(ANN)}
    \end{minipage}
    \begin{minipage}{0.16\hsize}
      \centering
      \captionsetup{justification=centering}
      \includegraphics[clip, width=\hsize]{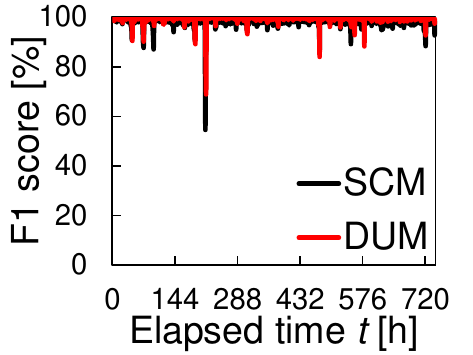}
      \subcaption{Australia East\\(GBDT)}
    \end{minipage}
    \begin{minipage}{0.16\hsize}
      \centering
      \captionsetup{justification=centering}
      \includegraphics[clip, width=\hsize]{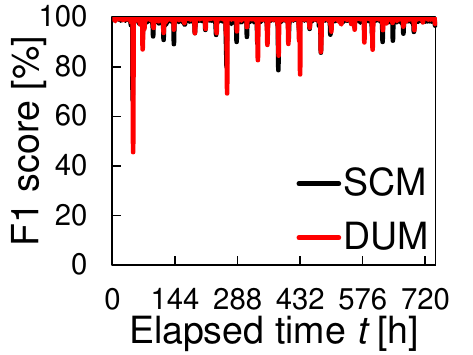}
      \subcaption{US East\\(GBDT)}
    \end{minipage}
    \begin{minipage}{0.16\hsize}
      \centering
      \captionsetup{justification=centering}
      \includegraphics[clip, width=\hsize]{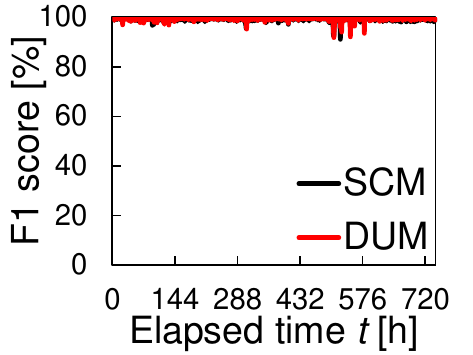}
      \subcaption{Japan East\\(GBDT)}
    \end{minipage}
    \vspace{1mm}
    \\
    \begin{minipage}{0.16\hsize}
      \centering
      \captionsetup{justification=centering}
      \includegraphics[clip, width=\hsize]{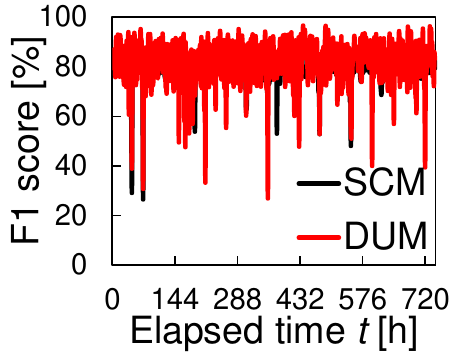}
      \subcaption{Australia East\\(TabNet)}
    \end{minipage}
    \begin{minipage}{0.16\hsize}
      \centering
      \captionsetup{justification=centering}
      \includegraphics[clip, width=\hsize]{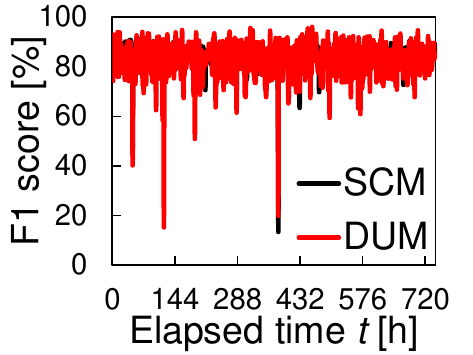}
      \subcaption{US East\\(TabNet)}
    \end{minipage}
    \begin{minipage}{0.16\hsize}
      \centering
      \captionsetup{justification=centering}
      \includegraphics[clip, width=\hsize]{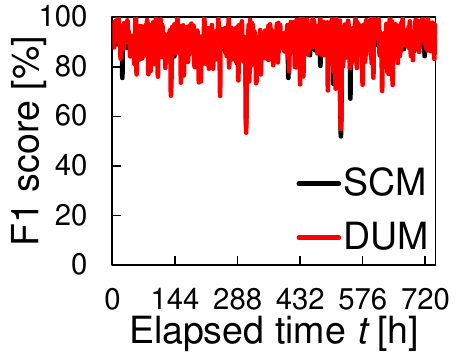}
      \subcaption{Japan East\\(TabNet)}
    \end{minipage}
  \end{tabular}
  \caption{Experiment 1: Transitions of F1 score ($T_{\mathrm{duration}}, T_{\mathrm{update}}=1$ [h]).}
  \label{f1-azure}
\end{figure*}

\subsection{Evaluation Indices}

Three indicators were adopted for evaluations: 
the distribution of destination ports of the malicious traffic observed by the honeypot server, the F1 score for the observed malicious traffic, and the training time for applying various machine learning algorithms.

First, we analyzed the distribution of destination ports of the observed traffic accessing the honeypot to identify the traffic and attack characteristics in each region. 
The distribution of destination ports was calculated on a per-flow basis. 
Note that we did not analyze it in Experiment 3 since it does not use the honeypot.

Next, we compared the evolution of the F1 score when traffic anomaly detection was performed using various machine learning algorithms for each anomaly detection model update method.
In all experiments, we confirmed the transitions of the F1 score when $T_{\mathrm{duration}}$ and $T_{\mathrm{update}}$ were set to one hour in SCM and DUM.

Additionally, we also evaluated $T_{\mathrm{duration}}$ and $T_{\mathrm{update}}$ were set to $\infty$ and one hour in DUM in Experiment 2, assuming that all captured traffic was permanently stored in real time.
This means that $T_{\mathrm{duration}}$ equals the time elapsed since the start of the experiment when this method is employed.
We compared the F1 scores for anomaly detection for the last hour of traffic in Experiment 2 by applying various machine learning algorithms to DUM with several different times $T_{\mathrm{duration}}$.

Furthermore, we measured the training time of various machine learning algorithms with several different $T_{\mathrm{duration}}$ in Experiment 2 to confirm the impact of the real-time model update on the training time. 
We used a computer with Intel Core i9-10980XE CPU and 256 GB RAM for training time measurement. 
The training process was executed by only using the CPU.

The machine learning algorithms used in the experiment were Support Vector Classification (SVC)~\cite{svc}, k-Nearest Neighbor (k-NN)~\cite{knn}, Decision Tree (DT)~\cite{dt}, Random Forest (RF)~\cite{rf}, Artificial Neural Network (ANN)~\cite{nn}, Gradient Boosting Decision Tree (GBDT)~\cite{gbdt}, and TabNet~\cite{tabnet}.
These algorithms except for TabNet were implemented by scikit-learn~\cite{scikit-learn}, and the TabNet algorithm was implemented by the pytorch-tabnet library~\cite{pytorch-tabnet}.
We used them with default parameters of the machine learning libraries. 
In the k-NN algorithm, the number of neighbors was set to 5.
In the RF algorithm, the number of trees was set to 100.
In the ANN algorithm, the number of hidden layers was set to 1, and the number of neurons was set to 100.
In addition, ReLU and Adam were adopted for the activation and optimization functions.
In the GBDT algorithm, the number of trees was set to 100.
In the TabNet algorithm, Adam was adopted as the optimization function.

\begin{figure*}[t!]
  \centering
  \begin{tabular}{cc}
    \begin{minipage}{0.2\hsize}
      \centering
      \includegraphics[clip, width=0.99\hsize]{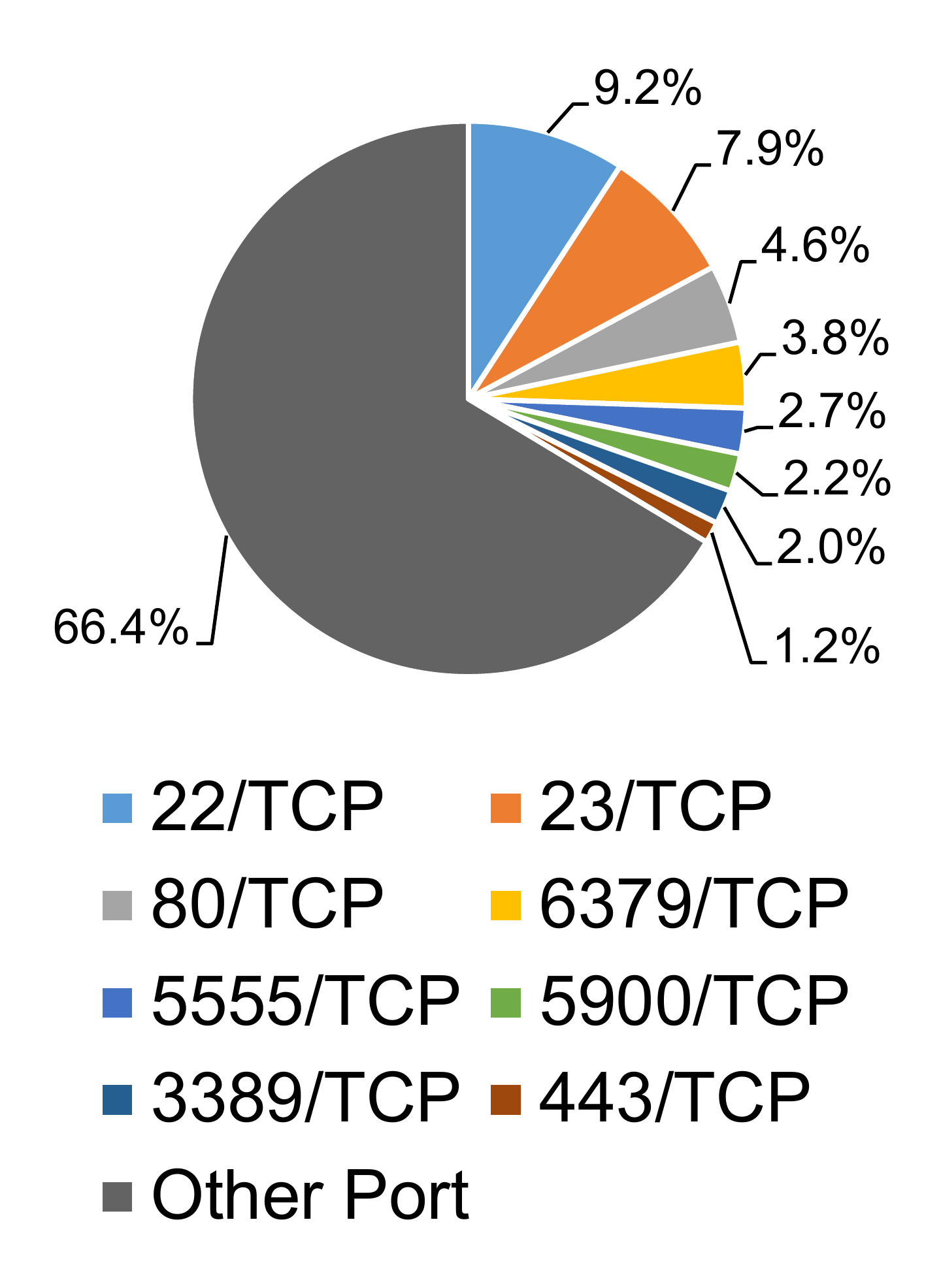}
	   \subcaption{Distribution of destination ports observed by the honeypot server.}
	   \label{exp2-a}
    \end{minipage}
    &    
    \begin{minipage}{0.8\hsize}
      \begin{tabular}{cccc}
        \begin{minipage}{0.23\hsize}
          \centering
          \includegraphics[clip, width=\hsize]{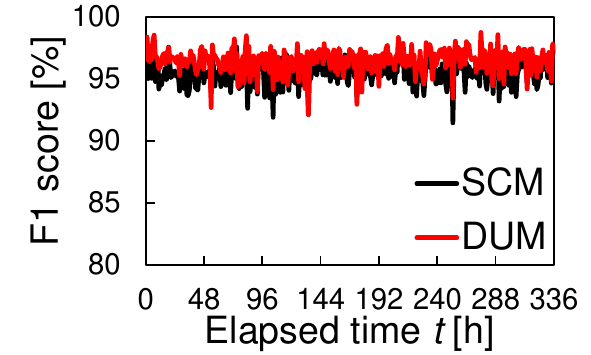}
          \footnotesize (a) SVC
        \end{minipage}
        \begin{minipage}{0.23\hsize}
          \centering
          \includegraphics[clip, width=\hsize]{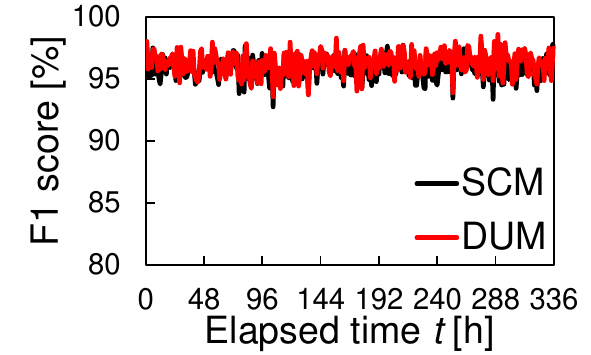}
          \footnotesize (b) k-NN
        \end{minipage}
        \begin{minipage}{0.23\hsize}
          \centering
          \includegraphics[clip, width=\hsize]{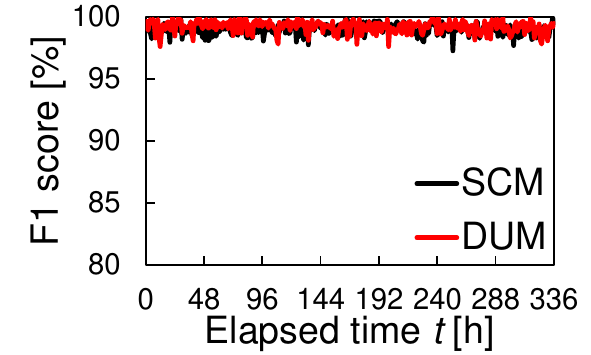}
          \footnotesize (c) DT
        \end{minipage}
        \begin{minipage}{0.23\hsize}
          \centering
          \includegraphics[clip, width=\hsize]{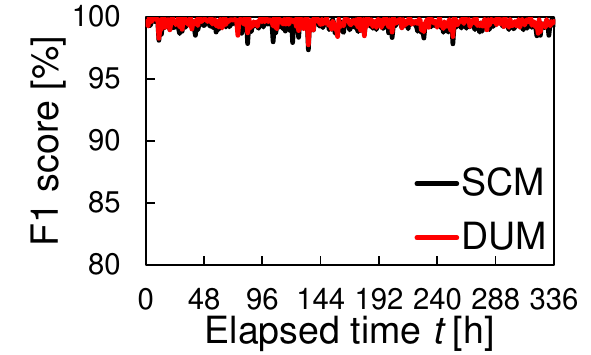}
          \footnotesize (d) RF
        \end{minipage}
      \\
        \begin{minipage}{0.23\hsize}
        \centering
        \includegraphics[clip, width=\hsize]{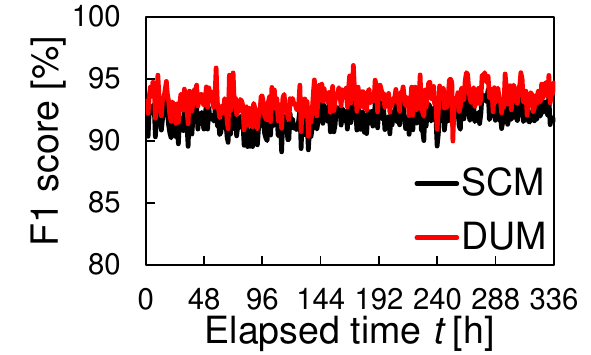}
        \footnotesize (e) ANN
      \end{minipage}
      \begin{minipage}{0.23\hsize}
        \centering
        \includegraphics[clip, width=\hsize]{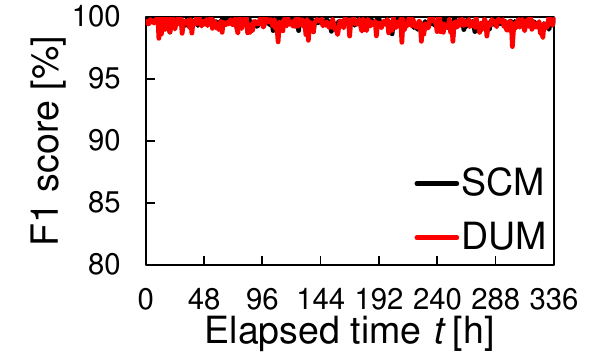}
        \footnotesize (f) GBDT
      \end{minipage}
      \begin{minipage}{0.23\hsize}
        \centering
        \includegraphics[clip, width=\hsize]{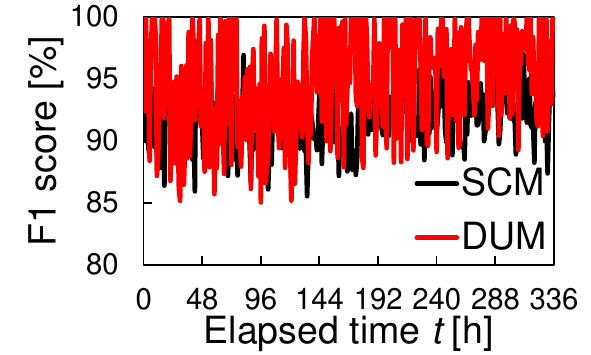}
        \footnotesize (g) TabNet
      \end{minipage}
    \end{tabular}
    \subcaption{Transitions of F1 score ($T_{\mathrm{duration}}, T_{\mathrm{update}}=1$ [h]).}
    \label{exp2-b}
    \end{minipage}
    \end{tabular}
  \caption{Experiment 2: Distribution of destination ports and transitions of F1 score.}
\end{figure*}

\begin{table*}[t!]  
  \centering
  \caption{Experiment 2: F1 scores for the last hour of traffic.}
  \label{tab:f1-changeT}
  \begin{tabular}{ccccccccc}
    \hline
      \diagbox{Algorithms}{$T_{\mathrm{duration}}$ [h]} & 1   & 2         & 4              & 12             & 24              & 48              & 168             & $\infty$       \\ \hline
      SVC                               & 97.70          & 98.03          & 96.28          & 95.93          & 96.28           & 98.77           & 98.77           & 98.37          \\
      k-NN                              & 97.53          & 98.21          & 96.96          & 96.68          & 96.98           & 99.48           & 99.83           & 96.26          \\
      DT                                & 99.52          & 99.68          & 99.68          & 99.52          & 99.68           & \textbf{100.0}  & 99.83           & 99.04          \\
      RF                                & \textbf{99.68} & \textbf{99.84} & \textbf{99.84} & 99.68          & \textbf{99.84}  & 99.95           & \textbf{100.0}  & \textbf{99.20} \\
      ANN                               & 94.70          & 94.86          & 94.44          & 97.87          & 97.19           & 98.77           & 99.13           & 94.53          \\
      GBDT                              & 99.52          & 99.68          & 99.76          & \textbf{99.76} & 99.68           & 99.48           & 99.84           & 99.04          \\
      TabNet                            & 99.32          & 99.48          & 98.77          & 99.13          & 98.37           & 98.77           & 98.77           & 98.21          \\ \hline
      Number of records for training    & 398            & 805            & 1588           & 4903           & 9722            & 19460           & 69307           & 135516         \\
    \hline
  \end{tabular}
\end{table*}

\section{Results and Discussion}
\label{result}

\subsection{Experiment 1}

Fig.~\ref{dstport-azure} shows a pie chart indicating the distribution of the destination ports with T-Pot in each region.
We observed port-scanning activities at remote access ports, such as 22/TCP and 3389/TCP, in all regions.
Additionally, we observed malicious traffic with infection-spreading activity by Mirai and its new variants on 23/TCP in all regions~\cite{mirai-1}. 
Therefore, cyberattacks targeting IoT devices were observed worldwide with varying patterns, which shows that we must continue handling such malicious traffic for self-defense.

Fig.~\ref{f1-azure} shows the transitions of the F1 score for various machine algorithms in each region, wherein it is evident that the proposed system can detect anomalous traffic. 
In particular, DUM detected anomalous traffic, including attacks, with higher accuracy than SCM when any machine learning algorithm was applied, despite the differences in the traffic and attack characteristics among regions.
These results prove that the proposed system can adapt to various attacks in real time by dynamically updating the anomaly detection model.

\subsection{Experiment 2}

Fig.~\ref{exp2-a} shows the distribution of destination ports observed by the honeypot server to analyze the traffic characteristics and attacks in an actual smart home environment.
We observed scanning activity by Mirai variants not only on 23/TCP, but also on 80/TCP and 5555/TCP, and vulnerable IoT devices infected with malware via these ports have been previously reported~\cite{port80,port5555}.
Therefore, as in Experiment 1, attackers continued to find and exploit vulnerabilities to perform infection-spreading activities similar to Mirai.

Fig.~\ref{exp2-b} shows the transitions of the F1 score when various machine learning algorithms were applied to the proposed system, to confirm its effectiveness in an actual smart home environment.
The results show that DUM could detect traffic anomalies with higher accuracy than SCM, similar to the results of Experiment 1.
These results confirm that the proposed system is self-adaptive to time-varying traffic, including anomalies, and can dynamically update the anomaly detection model.

Table~\ref{tab:f1-changeT} lists the F1 scores for the last hour of traffic in Experiment 2 in which several different $T_{\mathrm{duration}}$ values were set in the DUM and various machine learning algorithms were applied.
The results show that using longer traffic capture times to update detection models usually (but not always) improves detection accuracy.

Table~\ref{tab:time-changeT} lists the training time in Experiment 2, in which several different $T_{\mathrm{duration}}$ values were set and various machine learning algorithms were adopted. 
The results indicate that the training time required to update the model is sufficiently shorter than the traffic capture time, even if traffic with a longer duration is used for training. 
Therefore, considering the relationship between the traffic capture time and detection accuracy shown in Table~\ref{tab:f1-changeT}, the proposed system can immediately train and frequently update the model while maintaining a high detection accuracy by using a longer traffic capture duration for the training process.

\begin{table*}[!t]  
  \centering
  \caption{Experiment 2: Training time [s] in each $T_{\mathrm{duration}}$.}
  \label{tab:time-changeT}
  \begin{tabular}{cccccccccc}
    \hline
      \multicolumn{2}{c}{\diagbox{Algorithms}{$T_{\mathrm{duration}}$ [h]}}          & 1               & 2               & 4               & 12              & 24               & 48               & 168           \\ \hline
      \multirow{4}{*}{SVC}                            & Min                          & 0.0015          & 0.0043          & 0.0133          & 0.1200          & 0.4962           & 2.0471           & 29.4394       \\
                                                      & Max                          & 0.0070          & 0.0138          & 0.0363          & 0.2095          & 0.6496           & 2.6080           & 81.3703       \\
                                                      & Average                      & 0.0025          & 0.0065          & 0.0200          & 0.1466          & 0.5752           & 2.3012           & 31.0691       \\
                                                      & Variance                     & 0.0001          & 0.0001          & 0.0001          & 0.0002          & 0.0010           & 0.0145           & 15.5525        \\ \hline
      \multirow{4}{*}{k-NN}                           & Min                          & 0.0004          & 0.0007          & 0.0011          & 0.0037          & 0.0078           & 0.0182           & 0.0836        \\
                                                      & Max                          & 0.0013          & 0.0021          & 0.0034          & 0.0132          & 0.0175           & 0.0308           & 0.1289         \\
                                                      & Average                      & 0.0005          & 0.0008          & 0.0014          & 0.0043          & 0.0092           & 0.0202           & 0.0890         \\
                                                      & Variance                     & 0.0001          & 0.0001          & 0.0001          & 0.0001          & 0.0001           & 0.0001           & 0.0001          \\ \hline
      \multirow{4}{*}{DT}                             & Min                          & 0.0007          & 0.0012          & 0.0025          & 0.0075          & 0.0198           & 0.0453           & 0.2100         \\
                                                      & Max                          & 0.0034          & 0.0066          & 0.0105          & 0.0193          & 0.0382           & 0.0861           & 0.3211          \\
                                                      & Average                      & 0.0012          & 0.0017          & 0.0034          & 0.0109          & 0.0242           & 0.0553           & 0.2448         \\
                                                      & Variance                     & 0.0001          & 0.0001          & 0.0001          & 0.0001          & 0.0001           & 0.0001           & 0.0006          \\ \hline
      \multirow{4}{*}{RF}                             & Min                          & 0.0946          & 0.1040          & 0.1337          & 0.2639          & 0.4844           & 0.9513           & 4.2693           \\
                                                      & Max                          & 0.1358          & 0.1648          & 0.2063          & 0.3621          & 0.6113           & 1.1966           & 4.7571          \\
                                                      & Average                      & 0.1005          & 0.1132          & 0.1480          & 0.2910          & 0.5289           & 1.0487           & 4.4650           \\
                                                      & Variance                     & 0.0001          & 0.0001          & 0.0001          & 0.0002          & 0.0004           & 0.0020           & 0.0064           \\ \hline
      \multirow{4}{*}{ANN}                            & Min                          & 0.0176          & 0.0363          & 0.1173          & 0.2952          & 0.5301           & 1.0767           & 5.1498          \\
                                                      & Max                          & 0.4456          & 0.2865          & 0.5039          & 1.3893          & 3.1197           & 7.7973           & 43.2752          \\
                                                      & Average                      & 0.1139          & 0.1462          & 0.2493          & 0.6558          & 1.3885           & 3.0928           & 17.9145          \\
                                                      & Variance                     & 0.0046          & 0.0019          & 0.0053          & 0.0478          & 0.2310           & 1.4163           & 39.4680         \\ \hline
      \multirow{4}{*}{GBDT}                           & Min                          & 0.0702          & 0.1200          & 0.2070          & 0.5849          & 1.1642           & 2.4036           & 9.4871           \\
                                                      & Max                          & 0.3165          & 0.2100          & 0.3240          & 0.7404          & 1.3565           & 2.6984           & 9.8523           \\
                                                      & Average                      & 0.0998          & 0.1376          & 0.2360          & 0.6305          & 1.2428           & 2.5493           & 9.6774           \\
                                                      & Variance                     & 0.0007          & 0.0001          & 0.0002          & 0.0005          & 0.0014           & 0.0043           & 0.0073           \\ \hline
      \multirow{4}{*}{TabNet}                         & Min                          & 0.4255          & 0.8379          & 1.7823          & 6.0293          & 12.0369          & 24.7615          & 90.2335          \\
                                                      & Max                          & \textbf{0.7133} & \textbf{1.2494} & \textbf{2.4602} & \textbf{7.1699} & \textbf{13.8776} & \textbf{27.4693} & \textbf{94.6060}  \\
                                                      & Average                      & 0.4910          & 1.0262          & 2.1003          & 6.4962          & 12.8928          & 26.1914          & 92.7707          \\
                                                      & Variance                     & 0.0048          & 0.0061          & 0.0118          & 0.0547          & 0.1571           & 0.4800           & 1.1270           \\
    \hline
  \end{tabular}
\end{table*}

\begin{figure*}[t!]
  \centering
  \begin{tabular}{cccc}
    \begin{minipage}{0.24\hsize}
      \centering
      \includegraphics[clip, width=\hsize]{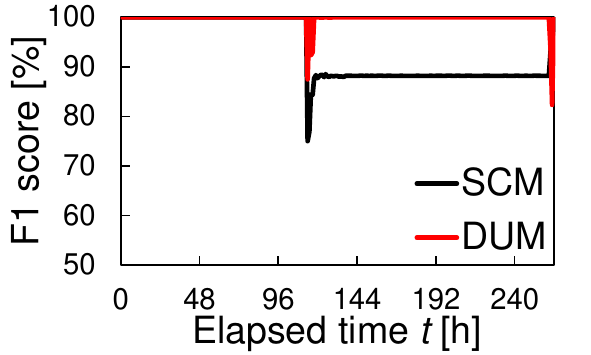}
      \subcaption{SVC}
    \end{minipage}
    \begin{minipage}{0.24\hsize}
      \centering
      \includegraphics[clip, width=\hsize]{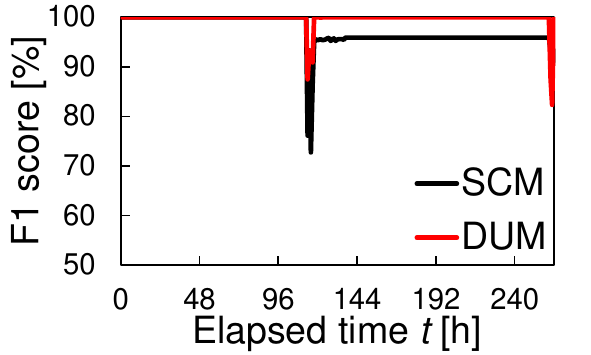}
      \subcaption{k-NN}
    \end{minipage}
    \begin{minipage}{0.24\hsize}
      \centering
      \includegraphics[clip, width=\hsize]{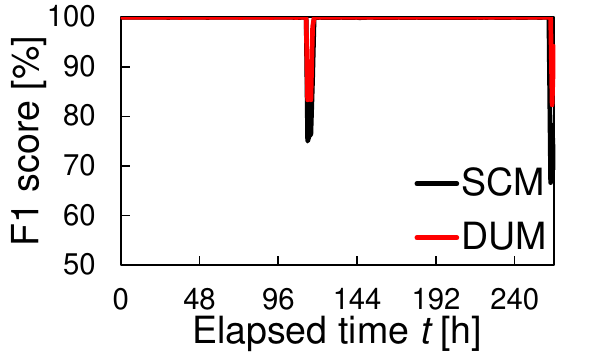}
      \subcaption{DT}
    \end{minipage}
    \begin{minipage}{0.24\hsize}
      \centering
      \includegraphics[clip, width=\hsize]{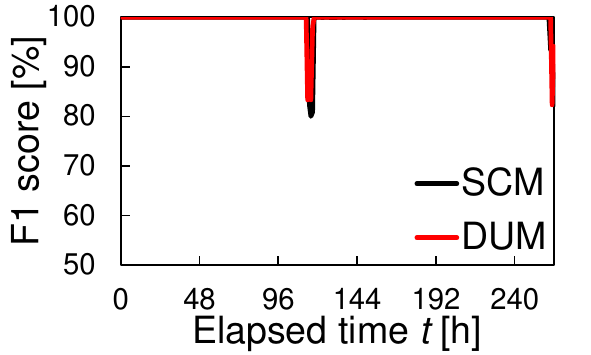}
      \subcaption{RF}
    \end{minipage}
  \\
  \vspace{3mm}
    \begin{minipage}{0.24\hsize}
      \centering
      \includegraphics[clip, width=\hsize]{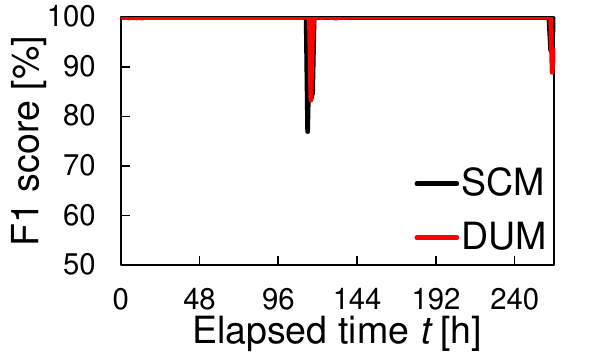}
      \subcaption{ANN}
    \end{minipage}
    \begin{minipage}{0.24\hsize}
      \centering
      \includegraphics[clip, width=\hsize]{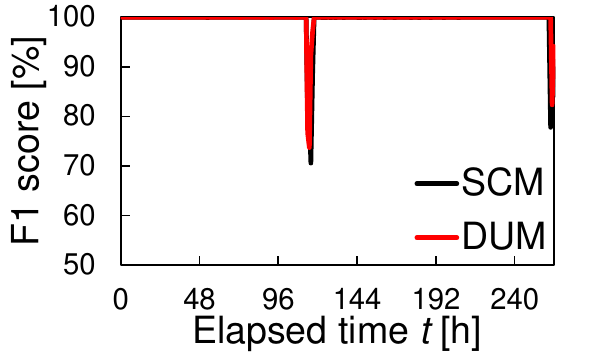}
      \subcaption{GBDT}
    \end{minipage}
    \begin{minipage}{0.24\hsize}
      \centering
      \includegraphics[clip, width=\hsize]{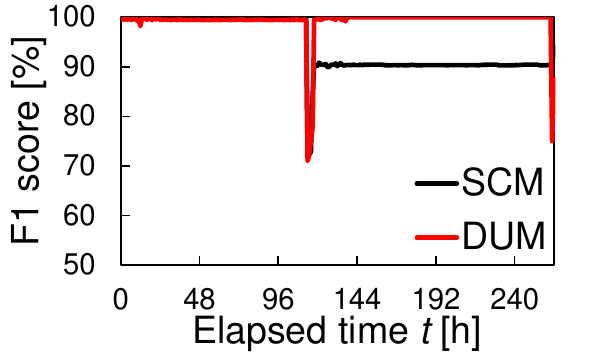}
      \subcaption{TabNet}
    \end{minipage}
  \end{tabular}
  \caption{Experiment 3: Transitions of F1 score ($T_{\mathrm{duration}}, T_{\mathrm{update}}=1$ [h]).}
  \label{f1-exp3}
\end{figure*}

\subsection{Experiment 3}

Fig.~\ref{f1-exp3} shows the transitions of the F1 score when using each machine learning algorithm in a mixed environment of normal and malware-infected abnormal devices. 
The results show that the F1 scores of SCM and DUM with any of the SVC algorithm, the k-NN algorithm, and the TabNet algorithm decreased about 120 hours after the start of the experiment. 
Afterwards, the F1 scores of DUM recovered. 
Therefore, these results prove that the abnormal traffic can be detected by updating the machine learning model in real time even if a local device is infected with a new malware.

\section{Conclusion}
\label{conclusion}

This paper proposed a self-adaptive anomaly detection system for IoT traffic in smart homes. 
The proposed system learns from real-time captured traffic and updates its model to dynamically adapt to anomalous traffic, including unknown attacks.
In addition, the proposed system is implementable as software modules on a real operating system, enabling the operator to use their preferred honeypot and gateway.

The experimental results demonstrated that dynamically updating the anomaly detection model using real-time captured traffic can improve the detection accuracy compared with the pre-generated detection model. 
The results of Experiment 1 indicated that the update method of the proposed system is effective pre-captured traffic from various regions of the world, despite the regional differences in the traffic characteristics. 
The results of Experiment 2 demonstrated that the proposed system works in an actual smart home environment by capturing real-time traffic and updating the detection model in real time.
The results of Experiment 3 proved that in an environment with a mixture of normal and malware-infected abnormal devices, it is possible to deal with changes in the behavior of the infected device by model updating.
These results indicate that IoT devices can be protected from various cyberattacks, including unknown attacks, by blocking communication with malicious hosts detected by the proposed system.

To examine the robustness of anomaly detection and its variation in the proposed system, long-term experiments at various points on the internet must be conducted because the proposed system is affected by the characteristics of the captured traffic and its temporal variation. 
Therefore, we plan to evaluate the proposed system under various traffic conditions to clarify its long-term performance as future work.

In addition, the detection results and reasons must be more interpretable to understand the characteristics of machine learning algorithms and model update in real time to further improve the detection accuracy. 
Interpretability allows the system to be adjustable because operators can find a strong dependency on specific features or a lack of important features for anomaly detection. 
For example, some information is expected to be used for improving detection accuracy: a port number that is often used for detecting abnormal communication and a period of time in which much traffic is generated.

Furthermore, we will consider a cooperative traffic anomaly detection system that simultaneously uses real-time traffic in multiple regions with cooperation among gateways, such as federated learning~\cite{fl-survey}, to immediately update the detection model according to region and improve its detection accuracy.

\section*{Acknowledgments}
The study results were partly obtained from the commissioned research (No. JPJ012368C05201) by National Institute of Information and Communications Technology (NICT), Japan.



\profile{Naoto Watanabe}{received the B.E. degree in electronic information systems from Shibaura Institute of Technology, Tokyo, Japan, in 2022. He is currently pursuing the M.S. degree at the Graduate School of Engineering and Science. His research interests include machine learning and cyber security for IoT.}

\profile{Taku Yamazaki}{received the B.E. and M.S. degrees in electronic information systems from Shibaura Institute of Technology, Tokyo, Japan, in 2012 and 2014, respectively. He received the D.E. degree in computer science and communications engineering from Waseda University, Tokyo, Japan, in 2017. He is presently an associate professor at Department of Electronic Information Systems, College of Systems Engineering and Science, Shibaura Institute of Technology, Saitama, Japan. His research interests include wireless networks, internet of things, and network security.}

\profile{Takumi Miyoshi}{received his B.Eng., M.Eng., and Ph.D.\ degrees in electronic engineering from the University of Tokyo, Japan, in 1994, 1996, and 1999, respectively. He started his career as a research associate in Waseda University from 1999 to 2001, and is presently a professor at Department of Electronic Information Systems, College of Systems Engineering and Science, Shibaura Institute of Technology, Saitama, Japan. He is also a research fellow in Institute of Industrial Science, the University of Tokyo, Tokyo, Japan. He was a visiting scholar in Laboratoire d'Informatique de Paris 6 (LIP6), Sorbonne Universit\'e, Paris, France, from 2010 to 2011. His research interests include overlay networks, location-based services, and mobile ad hoc and sensor networks.}

\profile{Ryo Yamamoto}{received his B.E.\ and M.E.\ degree in electronic information systems from Shibaura Institute of Technology, Tokyo, Japan, in 2007 and 2009. He received D.S.\ in global telecommunication studies from Waseda University, Tokyo, Japan, in 2013. He was a research associate at Graduate School of Global Information and Telecommunication Studies, Waseda University, from 2010 to 2014, and has been engaged in researching wireless communication networks. He is presently an associate professor at Graduate School of Informatics and Engineering, The University of Electro-Communications. He received the IEICE young researcher’s award in 2010, the IEICE Network System Research Award in 2014, the CANDAR/ASON Best Paper Award in 2014, IEICE Communications Society Distinguished Contributions Award in 2017, IEICE Information and Communication Management Distinguished Contributions Award in 2014 and 2017. His current research interests are ad hoc networks, sensor networks, IoT/M2M networks, and network protocols for the networks. He is a member of IEICE.}

\profile{Masataka Nakahara}{received the B.Eng. degree of Electrical Engineering and the M. Informatics degree of Graduate School of Informatics from Kyoto University, Japan in 2014 and 2016, respectively. He joined KDDI in 2016, and joined KDDI Research, Inc. in 2019. His current research interest includes cyber security for IoT.}

\profile{Norihiro Okui}{received the B.E. and M.E. degrees in Computer Science and Engineering from Waseda University, Japan, in 2010 and 2012, respectively. He joined KDDI in 2012. He is a research engineer at the Cyber Security Lab. in KDDI Research, Inc. His research interest includes cyber security for IoT.}

\profile{Ayumu Kubota}{received the B.E. and M.E degrees from Kyoto University, Japan, in 1993 and 1995, respectively. He joined KDD (now KDDI) in 1995, and has been engaged in the research on computer networks and cyber security.}

\end{document}